\documentclass[letter, twocolumn]{jpsj3} 
\usepackage{bm}

\title{Majorana Edge Modes of Superfluid $^3$He A-phase in a Slab}

\author{Yasumasa TSUTSUMI, Takeshi MIZUSHIMA, Masanori ICHIOKA, and Kazushige MACHIDA 
}
\inst{Department of Physics, Okayama University, Okayama 700-8530, Japan 
}
\abst{
Motivated by a recent experiment on the superfluid $^3$He A-phase with a chiral $p$-wave pairing confined in a thin slab, we propose designing a concrete experimental setup for observing the Majorana edge modes that appear around the circumference edge region. We solve the quasi-classical Eilenberger equation, which is quantitatively reliable, to evaluate several observables. To derive the property inherent to the Majorana edge state, the full quantum mechanical Bogoliubov-de Gennes equation is solved in this setting. On the basis of the results obtained, we perform decisive experiments to check the Majorana nature.
}

\kword{superfluid phase of $^3$He, restricted geometries, Majorana fermion, surface Andreev bound state}

\begin{document}
\maketitle

There has been growing interest in Majorana quasi-particles (QPs) and fermions,
which are expected to play a major role in various research fields,
ranging from cosmology and high-energy physics to condensed matter physics,~\cite{wilczek}
and even to topological quantum computing.~\cite{nayak}
The Majorana QP and fermionic operator are defined by $\gamma\!=\!\gamma^{\dagger}$ and $\Psi \!=\!\Psi^{\dag}$, respectively, 
which imply that the particle and antiparticle are identical  and 
thus are electrically neutral. 
They are an intriguing subject to further study in their own right.
It has been proposed that the Majorana nature brings new physics, such as the 
non-Abelian statistics for spinless chiral superfluids~\cite{nayak} and Ising-like spins for time-reversal invariant superfluids.~\cite{zhang,volovik,nagai}
Obviously, we like to have more candidate materials to realize it.

The zero-energy bound state appears whenever the underlying potential
for QP changes its sign. This is the so-called $\pi$-phase shift physics
exemplified by charge or spin density waves~\cite{takayama,fujita}, FFLO states~\cite{casalbuoni,nakanishi}, 
or stripes in high-$T_c$ cuprates~\cite{zaanen}. The Majorana QP is a special case of this type:
The Majorana conditions~\cite{read} under which the Majorana QP $\gamma^{\dagger}_{0}\!=\!\gamma_{0}$ is produced necessarily at a zero energy
are summarized as 
(1) chiral $p$-wave superconductors or superfluids where the Bogoliubov
QP $\gamma^{\dag}_E$ with an energy $E$ satisfies $\gamma^{\dagger}_E\!=\!\gamma_{-E}$
owing to particle-hole symmetry, and (2) a certain type of vortices with an odd winding number. For the edge or surface, the Majorana fermion $\Psi\!=\!\Psi^{\dag}$ exists as well as $\gamma^{\dagger}_{0}\!=\!\gamma_{0}$~\cite{read,stone,qi,qi2010,linder}.


There is as yet no firmly established chiral $p$-wave superconductors so far, 
although several theories~\cite{read,stone,furusaki} have explored its possibility in Sr$_2$RuO$_4$.
We have to resort to the use of the superfluid ${\rm ^3He}$ system in which we may 
manage to manipulate the order parameter (OP) so that the Majorana conditions
are fulfilled by setting up an appropriate boundary condition.
A good example of this kind has been recently presented by several groups~\cite{zhang,volovik,nagai}. 
They demonstrate that the Majorana edge modes in ${\rm ^3He}$-B give rise to the anisotropic spin dynamics~\cite{zhang,volovik,nagai}, while 
the OP in the B-phase is a time-reversal invariant, which is in contrast to that in the A-phase.
Thus the main purpose of the present work is to propose a concrete experimental
design for detecting the Majorana QP based on quantitatively accurate microscopic calculations
for the ${\rm ^3He}$ A-phase.

Here, we examine the superfluid ${\rm ^3He}$ A-phase, which is characterized by a chiral
$p$-wave type OP without doubt which is a time-reversal breaking state~\cite{salomaa}. Thus, 
the Majorana QPs in the A- and B-phases are worthy of comparative studies.
In fact, as we will see later, we can treat the Majorana QPs in the A- and B-phases in the same experimental setup
in a slab geometry by merely changing temperature or pressure, which is a large advantage over other proposed systems. 
The system provides a useful testing platform for checking various ideas associated
with the so-called topological order.

Our basic idea is stimulated by the experimental setup used 
by Bennett {\it et al.}~\cite{saunders}, where a superfluid ${\rm ^3He}$ A-phase  is confined in 
a thin slab with a thickness $D\!=\!0.6$ $\mu{\rm m}$ and  $(L\!=\!10{\rm mm}) \!\times \! 7{\rm mm}$, as schematically shown in Fig.~1. 
Other experimental groups~\cite{richerdson,xu, hata} use various-thickness samples, ranging from $D\!=\! 0.1$ to $10$ $\mu{\rm m}$.
Since the so-called dipole coherence length $\xi_d \!\sim\! 6$ $\mu{\rm m}$~\cite{salomaa},
the $l$-vector, which characterizes the chiral direction of the A-phase OP, points perpendicular to the
slab's upper and lower surfaces (B), as shown in Fig.~1. We assume the specular boundary 
condition in this paper. The four sides of the rectangular strips of the slab are an ideal place to accommodate
the Majorana edge modes.

We use two theoretical methods, the quasi-classical Eilenberger equation and Bogoliubov-de Gennes (BdG)
equation. The quasi-classical framework 
is used for detailed studies of the quantitative aspects
of the superfluid properties at the edges.  Using the BdG framework, we reveal the Majorana nature and quantum-mechanical spectral structures 
of the low-lying edge modes.

We consider the following spatially one-dimensional problem along the $x$-axis with a length $L$ ($0 \!\le\! x \!\le\! L$). 
See section A in Fig.~1 where, along the $z$-direction, the system is uniform because the ${\bm l}$-vector points 
perpendicular to the upper (B) and lower faces. The length is scaled by the coherence length at $T\!=\! 0$,
$\xi _0 \!\approx\! 80$ nm. In this situation, the pair potential of the chiral OP in the A-phase is described by
\begin{eqnarray}
\Delta(x, {\bm k})=\eta_x(x)k_x+\eta_y(x)k_y.
\label{eq:delta}
\end{eqnarray} 
Note that we are considering the three-dimensional Fermi sphere for $^3$He in the momentum space.
The relative momentum of the Cooper pair on the spherical Fermi surface is given by ${\bm k}\!=\!(\sin\theta_k\cos\phi_k,\sin\theta_k\sin\phi_k,\cos\theta_k)$,
and the Fermi velocity is given by ${\bm v}\!=\!v_{{\rm F}0}{\bm k}$.
It is important to notice that the quasi-particles described by Green's functions move under the ``given'' pair potential $\Delta(x,{\bm k})$.
The relative phase between $\eta_x$ and $\eta _y$  is $\pi/2$, since the ${\bm l}$-vector points to the $z$-direction.

\begin{figure}
\includegraphics[width=4cm]{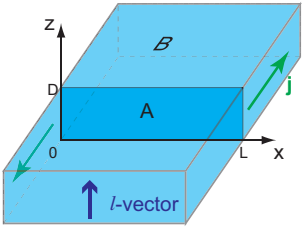}
\caption{(Color online) 
Schematic configuration of the whole system.
A: the cross-section considered here. B: Upper surface
of the container. The $l$-vector is locked to the
upper and lower surfaces. The mass current $j$ 
flows along the edges.
}
\end{figure}

The applied field ${\bm H}$ is an important controlling parameter of the system: For the slab thickness $D \!\leq\! \xi_d$,
the field perpendicular (parallel) to the slab should be $H \!>\! H_{d}$ ($H$ arbitrary) so as to satisfy the condition that ${\bm d} \!\perp\! {\bm H}$
with $H_d$ being a dipolar field $\sim\! 2{\rm mT}$~\cite{fetter}. 
Then, an equal spin pairing is realized and we can always ignore the $\uparrow\downarrow+\downarrow\uparrow$ Cooper pair by changing 
the spin quantization axis appropriately~\cite{kawakami}. Thus, we omit the spin degrees of freedom in our calculation, {\it e.g.}, in eq.~(\ref{eq:delta}).

It is known that different from that in the bulk ${\rm ^3He}$,  in the slab of superfluid ${\rm ^3He}$, 
the A-phase is stabilized at much lower temperatures $T$  down to the pressure $P\!=\!0$~\cite{ho}.
This condition is fulfilled for $0.06\!<\! D \!\lesssim \! 0.3$ $\mu{\rm m}$ 
where the A-phase is stable at $T\!=\!0$ and $P\!=\!0$. For sub-$\mu{\rm m}$ $D$ the A-phase changes
into the B-phase with decreasing $T$. Thus, it allows us to treat both phases under the same
experimental setup. A-B control is also possible by varying pressure.
Theoretically, our calculations are reliable quantitatively
because superfluid  ${\rm ^3He}$ in the lower-pressure regions of interest here can be described in a
weak coupling scheme without delicate strong coupling effects
\cite{greywall}.

We start with the quasi-classical Eilenberger equation~\cite{eilenberger,serene,ichiokaQCLs,ichiokaP,miranovic},
which has been used in the study of $^3{\rm He}$ 
superfluidity~\cite{schopohl,fogelstrom,sauls}. 
The quasiclassical Green's functions 
$g( \omega_n, {\bm k},x)$,  
$f( \omega_n, {\bm k},x)$, and  
$f^\dagger( \omega_n, {\bm k},x)$  
are calculated using the Eilenberger equation 
\begin{eqnarray} 
\left( \omega_n + \hat{\bm v} \cdot \nabla \right) f =\Delta g, \hspace{2mm}
\left( \omega_n - \hat{\bm v} \cdot \nabla \right) f^\dagger =\Delta^\ast g, 
\label{eq:Eil} 
\end{eqnarray}  
where $g \!=\! (1-ff^\dagger)^{1/2}$, ${\rm Re} g > 0$,  
and $\hat{\bm v} \!=\! {\bm v}/v_{{\rm F}0}$. 
We solve eq.~(\ref{eq:Eil}) by the Riccati method~\cite{miranovic,riccati1,riccati2}, with the specular boundary condition at both edges: $x\!=\! 0$ and $L$.
The temperature is measured using the transition temperature $T_{\rm c}$. Matsubara frequency $\omega _n \!=\! (2n+1)\pi k_{\rm B}T$, 
energy $E$, and pair potential $\Delta$ are in a unit of $\pi k_{\rm B}T_{\rm c}$.

The order parameter $\eta_j(x)$ ($j=x,y$)  
is self-consistently calculated by  
\begin{eqnarray} 
\eta_j(x) 
= 3g_0N_0 T \sum_{0 < \omega_n < \omega_{\rm cut}}  
 \left\langle k_j \left(  
    f +{f^\dagger}^\ast \right) \right\rangle_{\bm k},
\label{eq:scD}  
\end{eqnarray}  
with  
$(g_0N_0)^{-1}=  \ln T +2 T 
        \sum_{0 < \omega_n < \omega_{\rm cut}}|\omega_n|^{-1} $.  
$\langle \cdots \rangle_{\bm k}$ indicates the Fermi surface average,  
and $N_0$ is the density of states (DOS)  
at the Fermi energy in the normal state.  
We use $\omega_{\rm cut}\!=\!40 k_{\rm B}T_{\rm c}$. 
The temperature is fixed at $T\!=\!0.2T_c$ throughout the paper.

Using the obtained self-consistent solutions,  
the mass current is calculated using  
\begin{eqnarray}
{\bm j}(x) = (j_x,j_y,j_z) 
\propto T  \sum_{0 < \omega_n < \omega_{\rm cut}}  
\langle \hat{\bm v} {\rm Im} g   
\rangle_{\bm k}.  
\end{eqnarray}
When we calculate the quasiparticle states, 
we solve Eq.~(\ref{eq:Eil}) with  
$ {\rm i}\omega_n \!\rightarrow\! E + {\rm i} \delta$.  
We typically use $\delta=0.005$. 
The local density of states (LDOS) for quasiparticles normalized by $N_0$ is obtained as  
\begin{eqnarray} 
N(E,x)= \langle {\rm Re } 
\{ 
g( \omega_n, {\bm k},x) 
|_{i\omega_n \rightarrow E + i \delta} \}\rangle_{\bm k} .  
\end{eqnarray} 

\begin{figure}
\includegraphics[width=8.5cm]{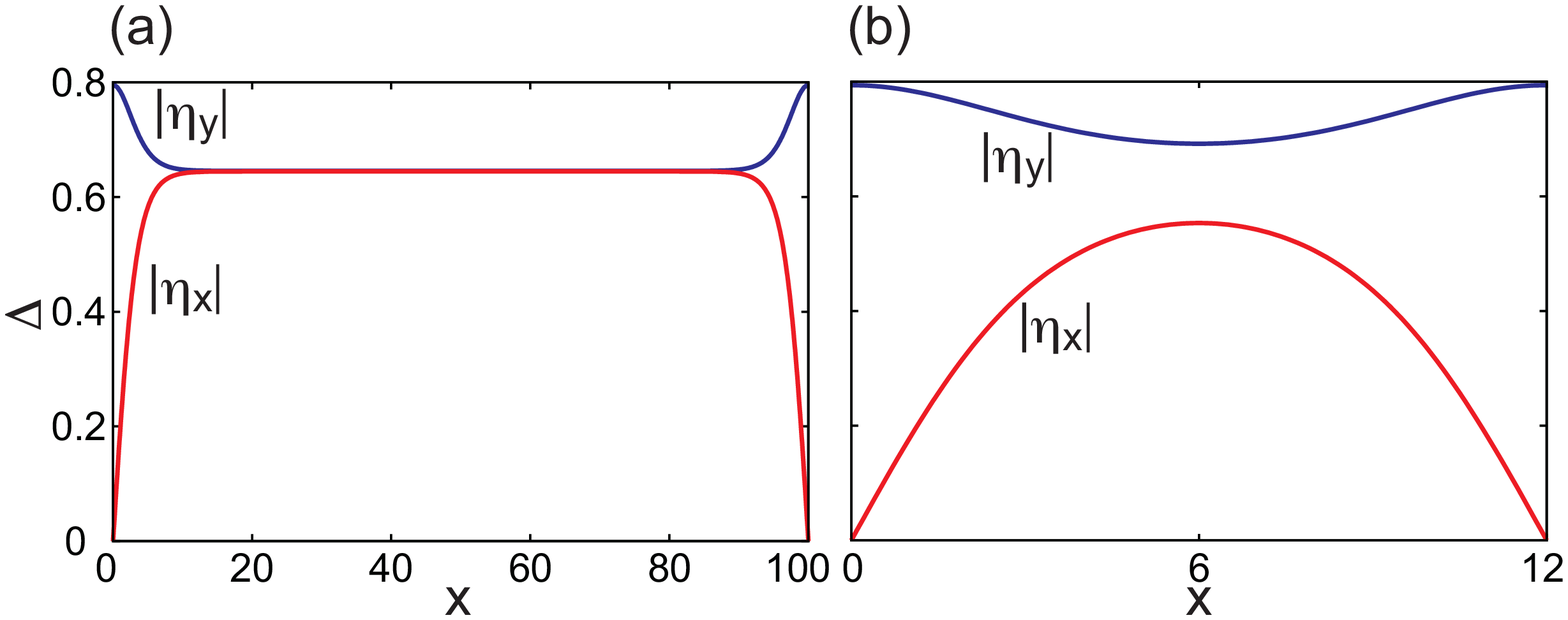}
\caption{(Color online) 
Order parameter profiles along the $x$-axis
for (a) $L=100$  and  (b) $L=12$ in a unit $\xi _0$.
}
\end{figure}

In Fig.~2(a), we show the OP profiles for $L\!=\!100$.
It is seen that, because of the specular boundary condition at the 
edges, the $k_x$ component, which is perpendicular to the edges,
becomes zero, while the parallel component $k_y$ is enhanced by
compensating for the loss of the $k_x$ component. Thus, 
at the edges, the polar state is realized. Towards the center, the
$k_x$ ($k_y$) component increases (decreases).
The A-phase with $k_x+ik_y$ is attained in the central region where the $l$-vector points to the $z$-direction.
We also show the result for $L\!=\!12$ in Fig.~2(b), 
where even in the central region the complete A-phase OP is not recovered
and the polar phase nature is  dominant there.

Figure~3(a) shows  LDOS at the edge $x\!=\!0$ and the center $x\!=\!50$.
It is clearly seen from curve 1 for $x\!=\!0$ that there exists a zero-energy state with a substantial weight
at $E\!=\!0$ corresponding to the Majorana edge mode, that is, LDOS is expressed as
$N(E, x\!=\!0)\!=\!\gamma_0+\alpha E^2$ near $E\!=\!0$.
The first (second) term comes from the Majorana edge mode (point node contribution
of the chiral state in the bulk A-phase). LDOS at $x\!=\!50$ (curve 2) shows a typical point node behavior of $N(E, x\!=\!50)\!\propto\! E^2$.

\begin{figure}
\includegraphics[width=8.5cm]{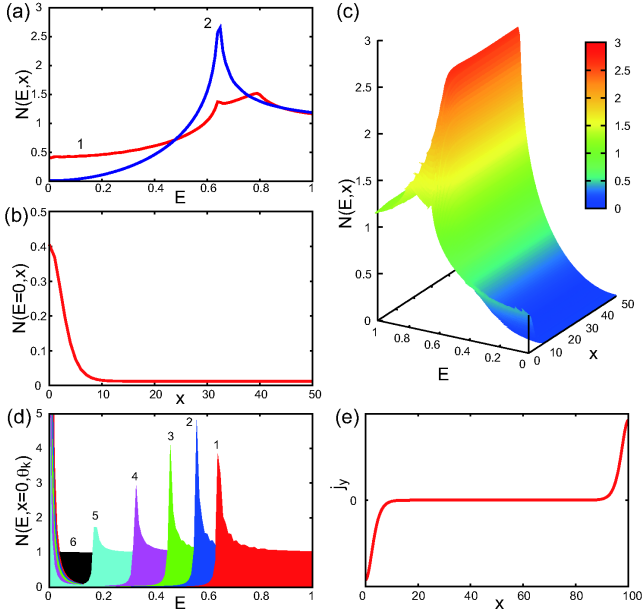}
\caption{(Color online) 
Calculated results for $L=100$.
(a) LDOS $N(E,x)$ at (1) $x=0$ and (2) $x=50$.
(b) Zero-energy LDOS $N(E\!=\!0,x)$ as a function of $x$ from the edge at $x\!=\!0$.
(c) Spectral evolution $N(E,x)$. The cross sections at $x\!=\!0$ and $x\!=\!50$ are same as in (a)
and $E\!=\!0$ as in (b). 
(d) The angle-resolved LDOS $N(E,x,\theta_k)$ in the momentum space at $x\!=\!0$  with 
$\phi_k \!=\! 0$ and $\theta_k \!=\! 89.5^{\circ}$ (1),
$60.5^{\circ}$ (2), $45.5^{\circ}$ (3), $30.5^{\circ}$ (4), $15.5^{\circ}$ (5), 
and $0.5^{\circ}$ (6). The zero-energy states are accumulated from all $\theta_k$ values, except for (6).
(e) Mass current $j_y(x)$.
}
\end{figure}

In Fig.~3(b), we show the extension of the Majorana edge mode towards the 
center from the edge at $x=0$, which spreads out an order of $\sim\!5\xi_0$.
In Fig.~3(c) the spectrum evolution from the edge to the center is shown.
The spectrum of the Majorana edge mode gradually changes into the bulk spectrum.
The angle-resolved LDOS $N(E,x,\theta_k)$ is shown in Fig.~3(d), where
 $\theta_k$ is the polar angle from the $k_z$-axis of the 3D Fermi sphere.
The zero-energy LDOS comes from the quasi-particles with the nonvanishing 
$k_x$ component, meaning that the QPs reflected by the edge form
the Andreev bound state exactly at $E\!=\!0$ because the QPs effectively feel 
the sign-reversed pair potential upon retracing the incoming path. 
The physics of the $\pi$-phase shift is working here.
The mass current $j_y(x)$ is displayed in Fig.~3(e), showing that 
the chiral current flows circularly along the four edges, as shown in Fig.~1.
By applying a magnetic field either parallel or perpendicular to the slab,
we can produce a spin imbalance due to the Zeeman shift 
between the $\uparrow\uparrow$ pairs and the $\downarrow\downarrow$ pairs.
This results in a net spin current in addition to the above mass current.

The LDOS's at the edge are calculated for various lengths in Fig.~4(a).
As $L$ decreases, the interference between two edges, which makes the zero energy levels split, 
becomes important and results in a decrease in the spectral weight of the zero-energy Majorana QP.
The spectral weight of the Majorana QP $N(E=0,x=0)$ is shown as a function of $L$ in Fig.~4(b).
Beyond $L\!\sim \! 20$, its weight remains constant.

\begin{figure}
\includegraphics[width=8.5cm]{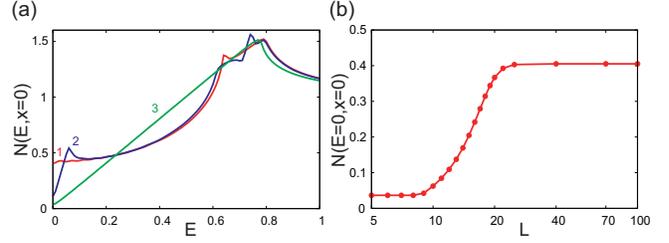}
\caption{(Color online) 
(a) LDOS at the edge $x\!=\!0$ for (1) $L\!=\!100$, (2) $L\!=\!12$, and (3) $L\!=\!8$.
(b) The spectral weight of the zero-energy state $N(E\!=\!0,x\!=\!0)$
as a function of $L$.
}
\end{figure}

To examine the full quantum nature of the discretized Majorana levels, we solve the BdG equation~\cite{mizushima08,mizushima10,mizushima10v2} with the pair potential 
obtained by the Eilenberger equation (\ref{eq:Eil}). Let us consider the situation with ${\bm d}\!\parallel\! \hat{z}$ for example. The BdG equation can be separated into two spin sectors labeled $\sigma \!=\! \pm$, which denotes the eigenstates of the Pauli matrix $\hat{\tau}_y$, corresponding to the choice of the spin quantization axis parallel to $\hat{y}$. Then the BdG equation reduces to each spin sector of the QP with the wave function ${\bm \varphi}_{n,\sigma} \!=\! [u_{n,\sigma},v_{n,\sigma}]^{\rm T}$ and the energy $E_{n,\sigma}$:~\cite{mizushima08,mizushima10,mizushima10v2}
\begin{eqnarray}
\int d{\bm r}_{2} 
\hat{\mathcal{H}}({\bm r}_1,{\bm r}_2) {\bm \varphi}_{n,\sigma}({\bm r}_2)
= E_{n} {\bm \varphi}_{n,\sigma}({\bm r}_1).
\label{eq:bdg}
\end{eqnarray}
Here, we set $(\hat{\mathcal H})_{11} \!=\! -(\hat{\mathcal{H}})_{22} \!=\! -\frac{\nabla^2}{2M}-E_{\rm F}$ and $(\hat{\mathcal{H}})_{12} \!=\! \Delta ({\bm r}_1,{\bm r}_2) \!=\! -(\hat{\mathcal{H}})^{\ast}_{21} $. The eigenstates of eq.~(\ref{eq:bdg}) yield one-to-one mapping between the positive energy states ${\bm \varphi}_E$ and the negative energy states ${\bm \varphi }_{-E} \!=\! \hat{\tau}_x{\bm \varphi}^{\ast}_E$ owing to the symmetry $\hat{\mathcal{H}} \!=\! -\hat{\tau}_x\hat{\mathcal{H}}^{\ast}\hat{\tau}_x$, leading to the relation of the Bogoliubov QP operator $\gamma _{E}\!=\! \gamma^{\dag}_{-E}$, as mentioned above.

Before turning to the numerical results of eq.~(\ref{eq:bdg}), we should mention the 
Majorana nature of the edge states. Within the weak coupling regime $k_{\rm F} \xi \!\gg\! 1$, eq.~(\ref{eq:bdg}) with eq.~(2) is solved for the edge states as ${\bm \varphi}_{{\bm k},\sigma} \!=\! e^{i\vartheta\hat{\tau}_z/2}e^{i{\bm K}\cdot{\bm r}_{\parallel}}f(k_x,x)$ with the energy $E \!=\! -\Delta _0 \frac{k_y}{k_{\rm F}}$, where ${\bm K} \!=\! (k_y,k_z)$, $f(k,x) \!=\! \sin{(k x)}e^{-x/\xi _0}$, and $\vartheta$ is the ${\rm U}(1)$ gauge of $\Delta$. Then, the field operator in the $y$-quantization axis is expanded in terms of the Bogoliubov operators $\gamma _{\bm K}$ as $\Psi _{\sigma} \!=\! \sum _{{\bm k}} [ e^{i{\bm K}\cdot{\bm r}_{\parallel}} \gamma _{\bm K} + {\rm h.c.}] f(k_x,x)e^{i\vartheta/2} + \Psi _{\rm bulk}$, where $\Psi _{\rm bulk}$ denotes the contribution from the bulk excitations. Although $\Psi _{\rm bulk}$ contains the low-energy states due to point nodes, their contributions are negligible if $T\!\ll\! T_c$ is considered. This is because the DOS due to the edge states yields $N(E) \!\propto\! E^{0}$ near $E\!\approx\! E_{\rm F}$, which overwhelms the DOS due to the point nodes $N (E)\!\propto\! E^2$, as shown in Figs.~3(a) and 3(b). This predicts the Majorana condition, $\Psi _{\pm} \!\approx\! e^{i\vartheta}\Psi^{\dag}_{\pm}$ for $T \!\ll\! T_c$ and $\gamma _{\bm K} \!=\! \gamma^{\dag}_{-{\bm K}}$.

In Figs.~5(a) and 5(b), we show the spectrum of the Majorana edge modes $E_n({\bm K})$
normalized by $\Delta_0$ at $T=0$. 
Here, the nonlocal pair potential in eq.~(\ref{eq:bdg}) is obtained by the self-consistent solution of the 
Eilenberger equation (\ref{eq:Eil}) through $\Delta ({\bm r}_1,{\bm r}_2) \!=\! \int \frac{d {\bm r}_{12}}{(2\pi)^3}
\Delta ({\bm r},{\bm k})e^{i{\bm k}\cdot {\bm r}_{12}}$, where ${\bm r}_{12} \!=\! {\bm r}_1 - {\bm r}_2$ is 
the relative coordinate.  
Along the $k_x$-axis, $E_n({\bm K} \!=\! 0)$ is dispersionless
and along the  $k_y$-axis two $k_y$-linear dispersions near $E\!=\!0$ appear, 
each coming from the left and right edges.
At $k_y\!=\!0$, the two Majorana modes are situated exactly at $E\!=\!0$.
A stereographic view of those modes is shown in Fig.~5(c).
It is seen that the linear Dirac dispersion continues along the $k_y$ direction,
forming a ``Dirac valley''.
The Dirac valley merges into point nodes at $k_z \!=\! \pm k_F$.
These results are consistent with those obtained using the previous Eilenberger
equation and reveal the detailed quantum level structures of Majorana modes.

\begin{figure}
\includegraphics[width=8.5cm]{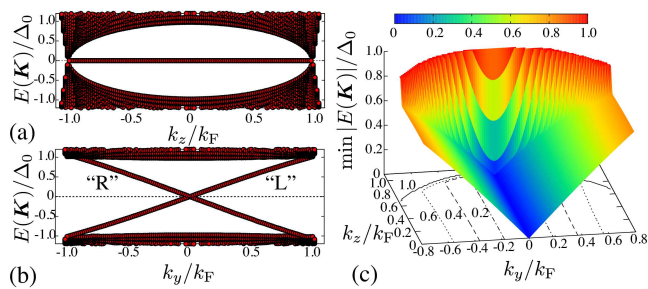}
\caption{(Color online)
Results of BdG calculations at $T\!=\! 0.2 T_{\rm c}$, where $\Delta (x,{\bm k})$ is taken from Fig.~2(a) with $L=100$. 
Eigenvalues of the low-lying excitations along (a) $k_z$ for $k_y \!=\! 0$  and 
 (b) $k_y$ for $k_z \!=\! 0$. ``R''(``L'') for right (left) edge. (c) 3D view of the dispersion relation:``Dirac valley''.
}
\end{figure}

There are several experimental ways to detect Majorana modes.
The most direct evidence of the Majorana nature is the observation of the anisotropy of spin susceptibility. Using the Majorana nature of the edge states $\Psi _{\pm} \!=\! e^{i\vartheta}\Psi^{\dag}_{\pm}$ for $T\!\ll\! T_c$, as derived above, the local spin operators result in $S_x \!\approx\! S_y \!\approx\! 0$ and only the $z$-component parallel to the ${\bm d}$-vector remains nontrivial, $S_z \!\approx\! -2i\Psi _+\Psi _-$. This predicts the Ising-like spin dynamics in the A-phase as well as in the B-phase \cite{zhang,nagai}. This is in sharp contrast to the anisotropy in the bulk A-phase, where the susceptibility parallel to ${\bm d}$ is suppressed at low $T$ according to the Yosida function.

The surface specific heat measurement, which was performed previously~\cite{bill}
in connection with the Andreev surface bound state detection in the B-phase,
resolves its contribution $C_{\rm surface}(T) \!=\! \gamma_{s}T$ because $\gamma_{s} \!\propto\! N(E\!=\! 0, x\!=\! 0)$
from the bulk contribution with $C_{\rm bulk} \!\propto\! T^3$, coming from the point nodes where $N(E) \!\propto\! E^2$.
Note that, in the slab geometry, there is no low-lying excitations other than $C \!\propto\! T^3$ 
from the two upper and lower surfaces (B in Fig.~1). Thus, the $C_{\rm surface}(T) \!=\! \gamma_{s}T$ contribution
of the Majorana QPs is distinctive.
The interference effect between Majorana QPs when $L$ decreases,
each coming from both edges, yields the LDOS $N(E) \!=\! N(0)+\alpha|E|$, as shown in Fig.~4.
This extra linear $|E|$ term gives the specific heat $C_{\rm surface}(T) \!\propto\! T^2$.
The relative weight of these $T$-linear and $T^2$ terms depends on the distance $L$
between the two edges, which is precisely evaluated (see for example $L \!=\! 12$ case (2) in Fig.~4(a)).

Quasi-particle scattering or QP beam experiments are extremely interesting.
They were performed in the past on $^4$He where roton-roton scattering is
treated~\cite{wyatt} and on the $^3$He B-phase where the Andreev surface 
bound state is investigated~\cite{okuda}. Using this method, we may pick up
Majorana QPs with a particular wave number because we like to 
manipulate Majorana QPs located at $k_y\!=\!0$, which is separated from 
other QPs in the nodal region at $k_z\!=\!\pm k_F$.

The other option might be to use a free surface where the Majorana surface 
bound state is formed.
As shown by Kono's group~\cite{kono}, the bound state can be detected through the 
excitation modes of the floating Wigner lattice of electrons placed on the surface.
We need a special, but feasible configuration of the experimental setups.


In summary, we have designed a concrete experimental setup  to observe the Majorana particles at the edge in a certain slab geometry. We calculate the microscopic Eilenberger equation to yield quantitatively reliable information on observable physical quantities in realistic situations for the $^3$He A-phase.
The Bogoliubov-de Gennes  equation is solved to explicitly demonstrate the Majorana nature of edge states. 
Several feasible and verifiable experiments for checking the Majorana nature are proposed.

The authors thank K. Kono and J. Saunders for informative discussions on their experiments.

\end{document}